\documentclass{aa}

\usepackage{lscape}
\usepackage{natbib}
\usepackage{graphicx}

\begin{document}

\title{The X-ray/SZ view of the virial region\\
\Large{II. Gas mass fraction}}

\author{D. Eckert\inst{1}  \and S. Ettori\inst{2,3} \and S. Molendi\inst{4} \and F. Vazza\inst{5,3} \and S. Paltani\inst{1}}
\institute{
Astronomical Observatory of the University of Geneva, ch. d'Ecogia 16, 1290 Versoix, Switzerland\\
\email{Dominique.Eckert@unige.ch}
\and
INAF - Osservatorio Astronomico di Bologna, Via Ranzani 1, 40127 Bologna, Italy
\and
INFN, Sezione di Bologna, viale Berti Pichat 6/2, 40127 Bologna, Italy
\and
INAF - IASF-Milano, Via E. Bassini 15, 20133 Milano, Italy
\and
Hamburg Observatory, Gojansbergsweg  112, 21029 Hamburg, Germany
}
\abstract{}{Several recent studies used the hot gas fraction of galaxy clusters as a standard ruler to constrain dark energy, which provides competitive results compared to other techniques. This method, however, relies on the assumption that the baryon fraction in clusters agrees with the cosmic value $\Omega_b/\Omega_m$, and does not differ from one system to another. We test this hypothesis by measuring the gas mass fraction over the entire cluster volume in a sample of local clusters.}{Combining the Sunyev-Zel'dovich thermal pressure from \emph{Planck} and the X-ray gas density from \emph{ROSAT}, we measured for the first time the average gas fraction ($f_{\rm gas}$) out to the virial radius and beyond in a large sample of clusters. We also obtained azimuthally-averaged measurements of the gas fraction for 18 individual systems, which we used to compute the scatter of $f_{\rm gas}$ around the mean value at different radii and its dependence on the cluster's temperature.}{The gas mass fraction increases with radius and reaches the cosmic baryon fraction close to $R_{200}$. At $R_{200}$, we measure $f_{{\rm gas},200}=0.176\pm0.009$ ($0.166 \pm 0.012$ for the subsample of 18 clusters in common between \emph{Planck} and \emph{ROSAT}). We find significant differences between the baryon fraction of relaxed, cool-core (CC) systems and unrelaxed, non-cool core (NCC) clusters in the outer regions. In average, the gas fraction in NCC clusters slightly exceeds the cosmic baryon fraction, while in CC systems the gas fraction converges to the expected value when accounting for the stellar content, without any evidence for variations from one system to another.}{We find that  $f_{\rm gas}$ estimates in NCC systems slightly disagree with the cosmic value approaching $R_{200}$. This result could be explained either by a violation of the assumption of hydrostatic equilibrium or by an inhomogeneous distribution of the gas mass. Conversely, cool-core clusters are found to provide reliable constraints on $f_{\rm gas}$ at overdensities $\Delta>200$, which makes them suitable for cosmological studies.}
\titlerunning{The X-ray/SZ view of the virial region II}
\maketitle

\section{Introduction}

The outskirts of galaxy clusters are the regions where the current activity of structure formation is taking place through the accretion of smaller structures onto massive clusters. In addition, they are the regions where the transition between virialized gas from clusters and accreting material from the large-scale structure occurs. Non-gravitational effects such as turbulence \citep[e.g.,][]{dolag05,vazza11a} and cosmic rays \citep[e.g.,][]{vazza12a,pfrommer07} are expected to play a more important role in these regions than in cluster cores, and the infall of smaller structures along large-scale filaments may cause the material in these regions to be clumpy \citep{nagai,simionescu} and asymmetric \citep{vazzascat,e12}. These effects may bias the measurements of cluster masses using X-ray and SZ proxies \citep{rasia,nagai07,piffaretti}, thus setting limitations on the use of galaxy clusters as high-precision cosmological probes \citep{allen11}. 

Baryons in galaxy clusters, mainly in the form of hot X-ray emitting plasma with some contribution from stars in member galaxies, are subjected to gravitational and radiative processes that, in a cold dark matter (CDM) scenario, affect their distribution in the DM halo accordingly. By forming in the highest peaks of the primordial gravitational fluctuations, massive galaxy clusters are relatively well-isolated gravity-dominated structures, suggesting that their mass function and relative baryon budget are highly sensitive tests of the geometry and matter content of the Universe. In particular, estimates of the gas and total mass content of relaxed hot clusters in hydrostatic equilibrium can be obtained using X-ray observations, allowing one both to place a lower limit on the cluster baryon mass fraction, which is expected to match the cosmic value $\Omega_{\rm b} / \Omega_{\rm m}$ \citep[e.g.,][]{white93}, and to constrain the cosmic dark energy by imposing the gas mass fraction as standard ruler \citep[see e.g.][]{ettori03,ettori09,allen08}.

In this paper, we combine the SZ pressure profiles from the sample of 62 galaxy clusters observed with the \emph{Planck} satellite \citep[hereafter P12]{planck5} with the \emph{ROSAT} gas density profiles presented in \citet[hereafter E12]{e12} to reconstruct gravitating mass profiles assuming hydrostatic equilibrium. This allows us to provide the first measurements of the gas fraction in cluster outskirts on relatively large samples, probing a volume several times larger than the typical \emph{XMM} and \emph{Chandra} limits and the entire azimuth (as opposed to most of the \emph{Suzaku} results). We study the importance of additional effects (non-thermal pressure support, gas clumping, asymmetry) on the quantities recovered by assuming that the gas is in hydrostatic equilibrium. Similar results are also presented for a subset of 18 individual clusters (6 CC and 12 NCC) that are in common between the samples of P12 and E12. In a companion paper (hereafter Paper I), we demonstrate the validity of the method and apply it to infer the average thermodynamic properties of galaxy clusters.

Throughout the paper, we assume a $\Lambda$CDM cosmology with $\Omega_m=0.3$, $\Omega_\Lambda=0.7$, and $H_0=70$ km s$^{-1}$ Mpc$^{-1}$. We also refer to the best-fit results on the $\Lambda$CDM model provided from WMAP-7 years \citep[see e.g. Table~8 in ][]{wmap7}, which imply a cosmic baryon fraction $\Omega_{\rm b} / \Omega_{\rm m} = 0.167 \pm 0.011$. Because our cluster sample is located at redshift below 0.2, the differences between these cosmological models affect the estimates of the gas mass fraction presented in this work by less than 1 per cent, through the variation of the angular diameter distance.

\begin{figure}
\resizebox{\hsize}{!}{\includegraphics{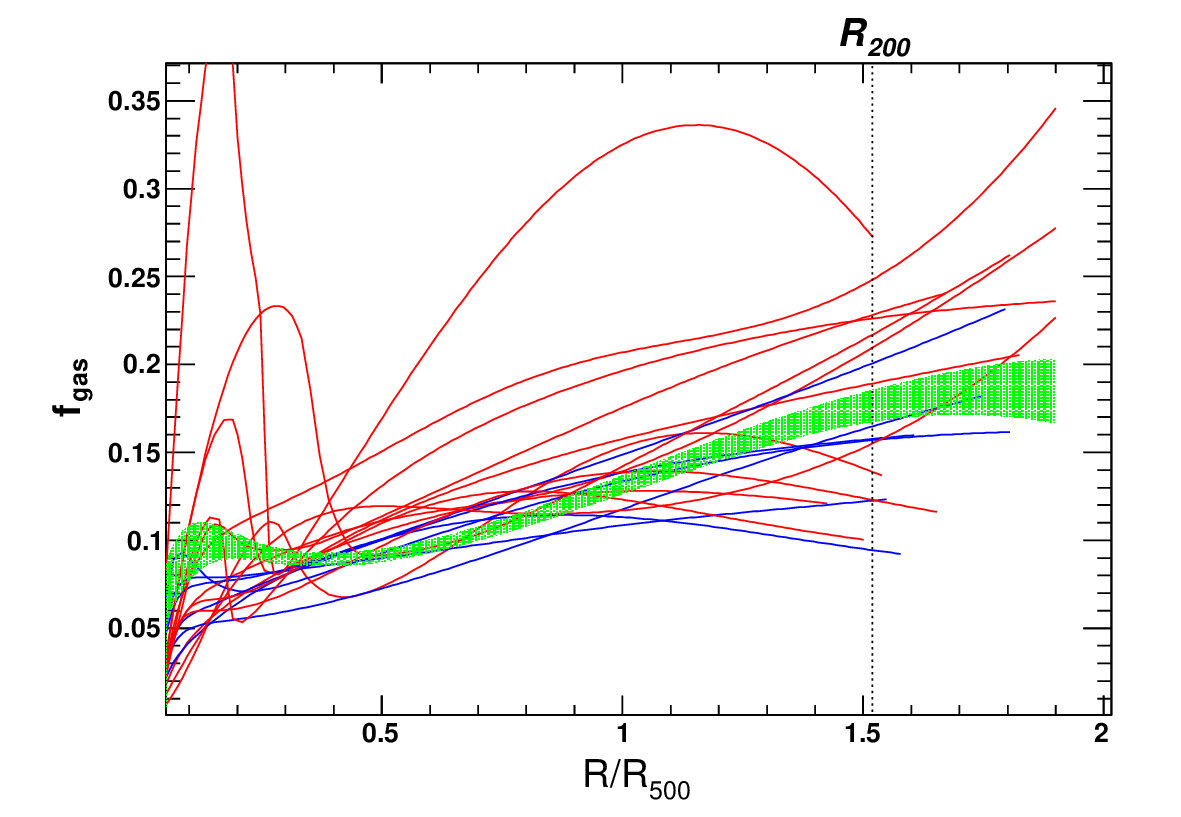}}
\caption{Observed gas mass fraction profiles for CC (blue) and NCC (red) clusters. All profiles stop at the maximum detection radius ($2 \sigma$ above the background) in ROSAT/{\it PSPC}. The green-shaded region shows the $f_{\rm gas}$ profile from the average gas density and {\it Planck} pressure profiles. The most strongly deviating profile is that of A2163.}
\label{fgas_all}
\end{figure}

\section{Basic formalism}

For this work, the two main observables are the pressure from SZ measurements and the gas number density from X-ray observations. In this case, the hydrostatic mass is simply given by 

\begin{equation} M_{{\rm hyd}}(<r)=-\frac{r^2}{G \mu m_{\rm p} n_{\rm gas}}\frac{{\rm d}P_{\rm gas}}{{\rm d}r}, \label{hydmass}\end{equation}

\noindent where $\mu=0.6$ is the mean molecular weight, while the gas mass profiles can be computed directly by integrating the density profiles,

\begin{equation} M_{\rm gas}(<r) = \mu m_{\rm p} \int_0^r n_{\rm gas}(r^\prime) 4\pi r^{\prime 2}\, dr^\prime . \label{gasmass}\end{equation}

\noindent An estimate of $f_{\rm gas}(r)=M_{\rm gas}(<r)/M_{\rm hyd}(<r)$ can thus be obtained in a straightforward way. To describe the pressure and the gas density, we fit the data point using general parametric functions (see paper I for details). Following P12, the pressure is described by a generalized Navarro-Frenk-White profile \citep[GNFW,][]{gnfw,nagai07b}. The pressure gradient can be expressed analytically by differentiating the GNFW profile. The pressure profiles are taken from P12, for the average population as well as for the 18 individual clusters in common between the two samples. 

For the density, we used a simplified version of the functional form introduced by \citet{vikhlinin06}. This functional form was projected along the line-of-sight and fit to the \emph{ROSAT} emission-measure profiles. The error budget was estimated through a Monte-Carlo Markov chain (MCMC) method, which allows us to compute the envelopes of the curves derived from the fitting procedure. Alternatively, we also used deprojected density profiles as computed in E12, and interpolated them along a common grid through a cubic spline method. For more details and a thorough validation of the method, we refer the reader to Paper I, also for the details of the error calculation. In Fig. \ref{fgas_all} we show the gas fraction profiles derived for the 18 individual systems and the average pressure and density profiles. Since we focus our work on the outer regions, the MCMC procedure was performed only on the data beyond $0.2R_{500}$, which causes the bumps at low radii. Beyond $0.5R_{500}$, the most strongly deviating profile is that of A2163, a violently merging cluster that is likely out of hydrostatic equilibrium (see Appendix C of Paper I for a discussion of this system).

\begin{figure*}
\hbox{
  \resizebox{0.5\hsize}{!}{\includegraphics{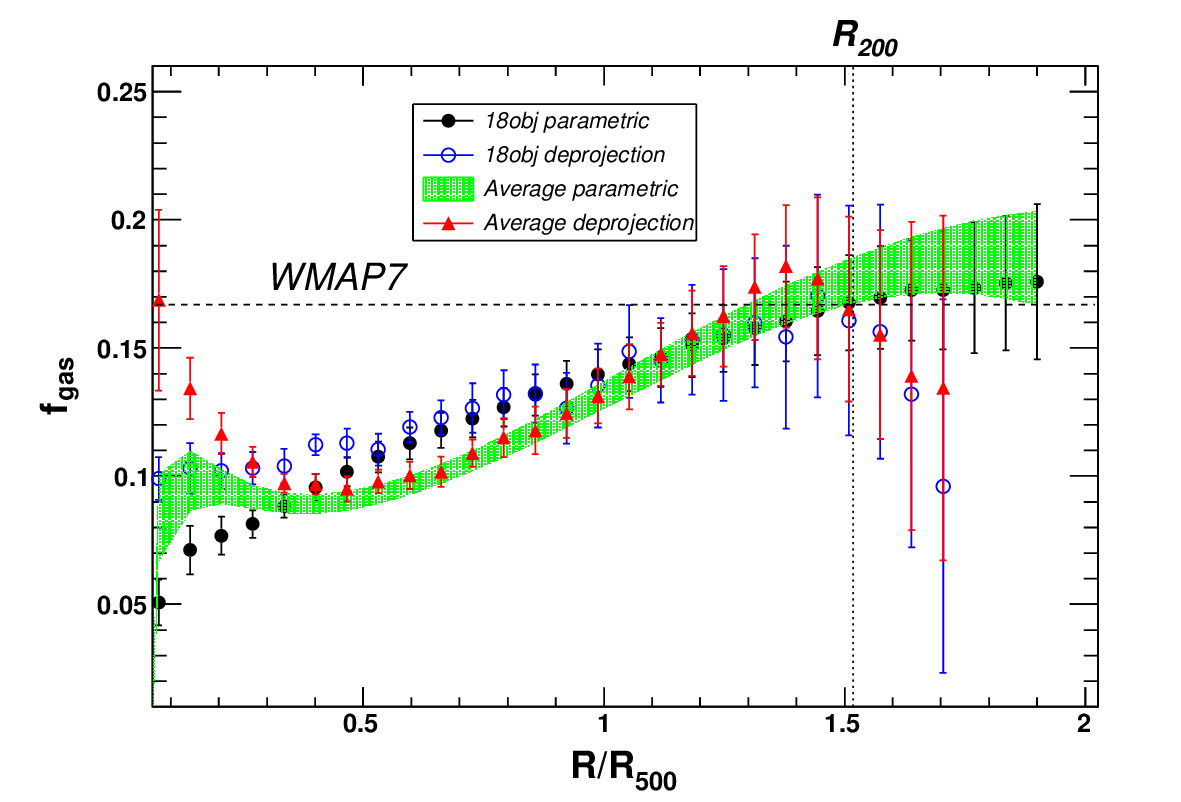}}
  \resizebox{0.5\hsize}{!}{\includegraphics{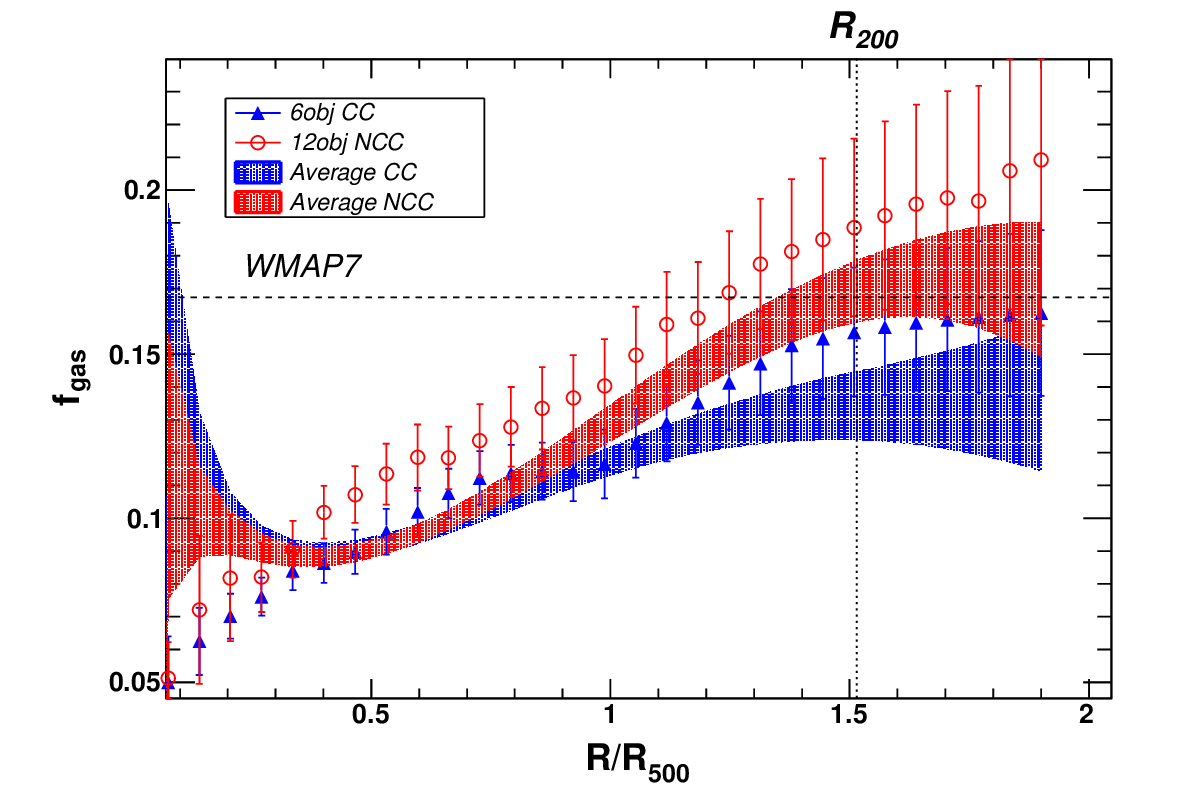}}
}
\caption{{\it (Left)} Average gas fraction profile computed from the sample-average pressure and density profiles and from the subsample of 18 individual objects. We show the profiles obtained through parametric forward fitting with MCMC error envelopes for both cases (sample average, green shaded area; median of 18 objects, black circles), and by interpolating the deprojected density profiles (sample average, red triangles; median of 18 objects, blue circles). The error bars are given at the $1\sigma$ level. 
({\it Right}) Average gas fraction profiles for the CC (blue) and NCC (red) populations separately. The shaded areas represent the profiles obtained from the sample-averaged profiles, while the data points show the median of the 18 individual systems. }
\label{fig:fgas}
\end{figure*}

\section{Results}

For a fiducial cluster with a virial temperature $kT=5$ keV, we used the universal pressure profiles from P12 together with the median gas density profiles from E12 to extract gas fraction profiles using the hydrostatic equilibrium equation (Eq. \ref{hydmass}). Unlike the gas fraction profiles presented in Sect. 7.3 of P12, no extrapolation is needed to compute the gas fraction. In Fig. \ref{fig:fgas}, we show the average gas fraction profile obtained by combining the average profiles from P12 and E12, computed through the parametric forward-fitting method (green-shaded area) and the geometrical deprojection method (red points). Excellent agreement is found between the results obtained between the two methods, and hence the results presented here are stable against the assumed deprojection technique. The profiles obtained from the sample-averaged profiles are also compared to the median of the 18 individual objects for both deprojection methods. The uncertainties in the median were estimated using 1000 bootstrap reshufflings of the sample (see Paper I). Below $1.7R_{500}$\footnote{For a given overdensity $\Delta$, $R_\Delta$ is the radius for which $M_\Delta/(4/3\pi R_\Delta^3)=\Delta\rho_c$}, all methods show a remarkable level of agreement, demonstrating the reliability of the measurements. In the following and for the sake of simplicity we focus on the results obtained with the parametric form, as in Paper I. However, we stress that the geometrical deprojection method always leads to consistent results.

At $R_{500}$, we find $f_{{\rm gas}, 500}=0.132\pm0.005$, which agrees well with previous estimates from {\it Chandra} \citep[e.g.,][]{vikhlinin06} and {\it XMM-Newton} \citep[e.g.,][]{ettori10}. The measured gas fraction reaches the cosmic baryon fraction \citep[$\Omega_b/\Omega_m=0.167$,][]{wmap7} at $R=1.4R_{500}$ and slightly exceeds it beyond this radius, although it remains consistent within the error bars. At $R_{200}$, we measure $0.176\pm0.009$ ($0.160\pm0.019$ in the subsample of 18 objects) with the forward-fitting method,  $0.158\pm0.019$ ($0.160\pm0.028$) with the deprojection technique.
Given that the stellar mass contributes another $10-20\%$ of the baryon fraction \citep[e.g.,][]{gonzalez,lagana}, the gas fraction recovered assuming hydrostatic equilibrium exceeds the expected value in cluster outskirts at the $2\sigma$ level. Although the confidence level is low, this may indicate that either the mass recovered through hydrostatic equilibrium is biased low, as suggested from present hydrodynamical numerical simulations \citep[e.g.,][]{rasia06,rasia12,nelson,burns10}, or that inhomogeneities in the gas distribution bias high the estimated $M_{\rm gas}$ \citep{nagai,simionescu}, or a combination of the two effects.

\begin{table*}
\caption{\label{tabscat}Mean $f_{\rm gas}$ values and intrinsic scatter $\sigma_f$ (in precent) obtained from various datasets and at different overdensity radii. The datasets labeled ``T-scaled" were rescaled by the quantity $(T_{vir}/\mbox{7 keV})^\beta$ (see Eq. \ref{funcover} and Table \ref{tableover}). Uncertainties are given at the $1\sigma$ confidence level, while upper limits are 90\%.}
\begin{center}
\begin{tabular}{lcccccccc}
\hline
Dataset & $f_{{\rm gas},2500}$ & $\sigma_{f,2500}$ & $f_{{\rm gas},1000}$ & $\sigma_{f,1000}$ & $f_{{\rm gas},500}$ & $\sigma_{f,500}$ & $f_{{\rm gas},200}$ & $\sigma_{f,200}$ \\
\hline
\hline
18 obj & $10.1\pm0.5$ & $1.5\pm0.4$ & $12.7\pm0.7$ & $2.4_{-0.6}^{+0.7}$ & $14.9\pm1.0$ & $3.7_{-0.8}^{+1.0}$ & $16.6\pm1.2$ & $4.2_{-0.9}^{+1.1}$\\
6 CC obj & $8.7\pm0.5$ & $0.9\pm0.4$ & $10.7\pm0.5$ & $0.6\pm0.6$ & $12.3\pm0.6$ & $<1.9$ & $14.3_{-1.4}^{+1.5}$ & $2.8_{-1.0}^{+1.4}$\\
12 NCC obj & $10.9\pm0.5$ & $1.2\pm0.6$ & $13.8\pm2.8$ & $2.8_{-0.8}^{+1.1}$ & $16.3\pm1.4$ & $4.4_{-1.0}^{+1.4}$ & $17.8\pm1.6$ & $4.7_{-1.1}^{+1.5}$\\
\\
18 obj, T-scaled & $9.9\pm0.3$ & $0.9\pm0.4$ & $12.6\pm0.5$ & $1.3\pm0.5$ & $14.6\pm0.7$ & $2.3_{-0.6}^{+0.7}$ & $16.4\pm1.1$ & $3.7_{-0.8}^{+1.0}$\\
 6 CC obj, T-scaled & $9.0\pm0.4$ & $<0.8$ & $11.3\pm0.5$ & $<1.0$ & $12.9\pm0.6$ & $<1.3$ & $14.9_{-1.5}^{+1.6}$ & $2.8_{-1.0}^{+1.5}$\\
12 NCC obj, T-scaled & $10.9\pm0.5$ & $1.9_{-0.5}^{+0.6}$ & $13.4\pm0.7$ & $1.6_{-0.6}^{+0.7}$ & $15.6\pm1.0$ & $2.8_{-0.7}^{+0.9}$ & $17.2_{-1.4}^{+1.5}$ & $4.0_{-1.0}^{+1.4}$\\
\hline
\end{tabular}
\end{center}
\end{table*}

Similar to the entropy (see Paper I), we observe differences in the gas fraction profiles computed for the 6 CC and 12 NCC clusters independently (see Fig. \ref{fig:fgas}). The NCC profiles tend to increase more steeply in cluster outskirts compared to CC profiles. From the sample-averaged profiles, at $R_{200}$ NCC clusters slightly exceed the cosmic baryon fraction ($f_{{\rm gas},NCC}=0.169\pm0.010$ using the average profiles; $0.178\pm0.016$ for the 12 individual NCC systems), while the gas fraction in CC systems is close to the expected gas fraction around $R_{200}$ ($f_{{\rm gas},CC}=0.134\pm0.011$; $0.143 \pm 0.015$ for the subsample of 6 CC objects). Despite its low statistical significance (less than 2$\sigma$), this result likely indicates that hydrostatic equilibrium breaks down in the outskirts of NCC systems, while in CC (relaxed) clusters the behavior is closer to self-similar expectations, in agreement with our results for the entropy (see Paper I). 

\begin{figure*}
\hbox{
\resizebox{0.5\hsize}{!}{\includegraphics{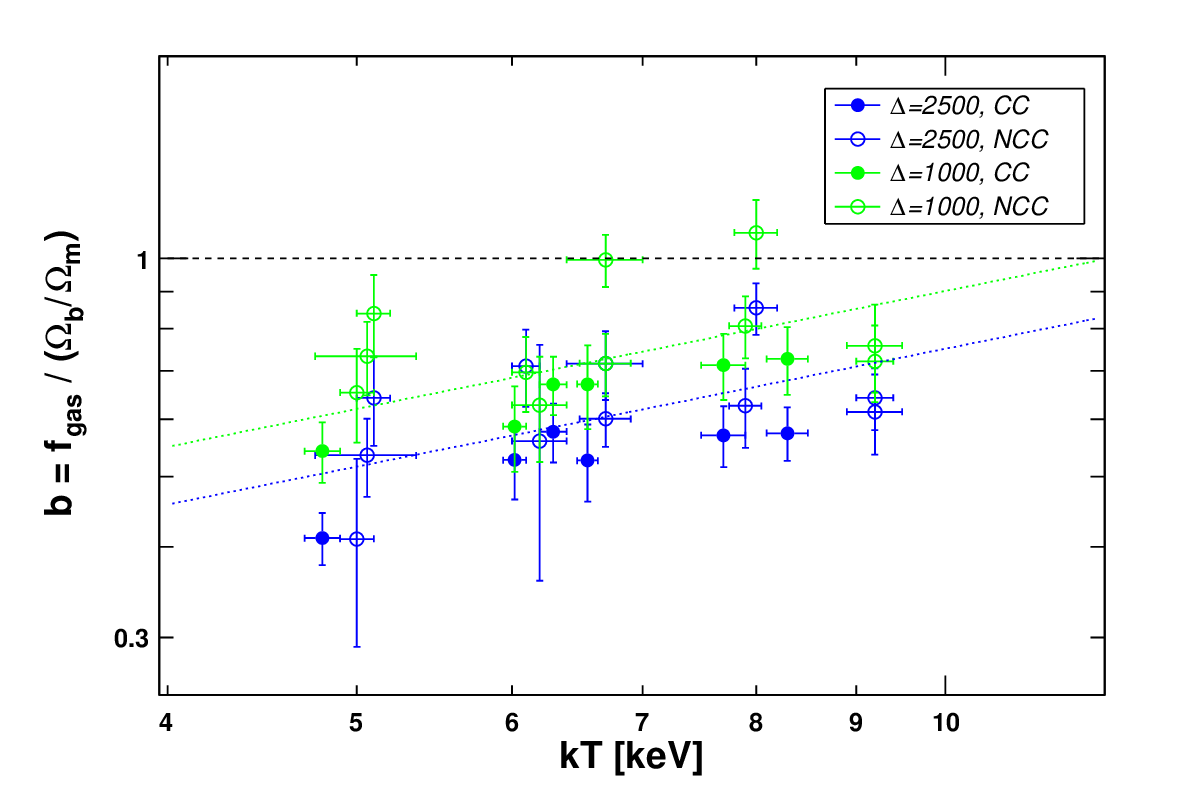}}
\resizebox{0.5\hsize}{!}{\includegraphics{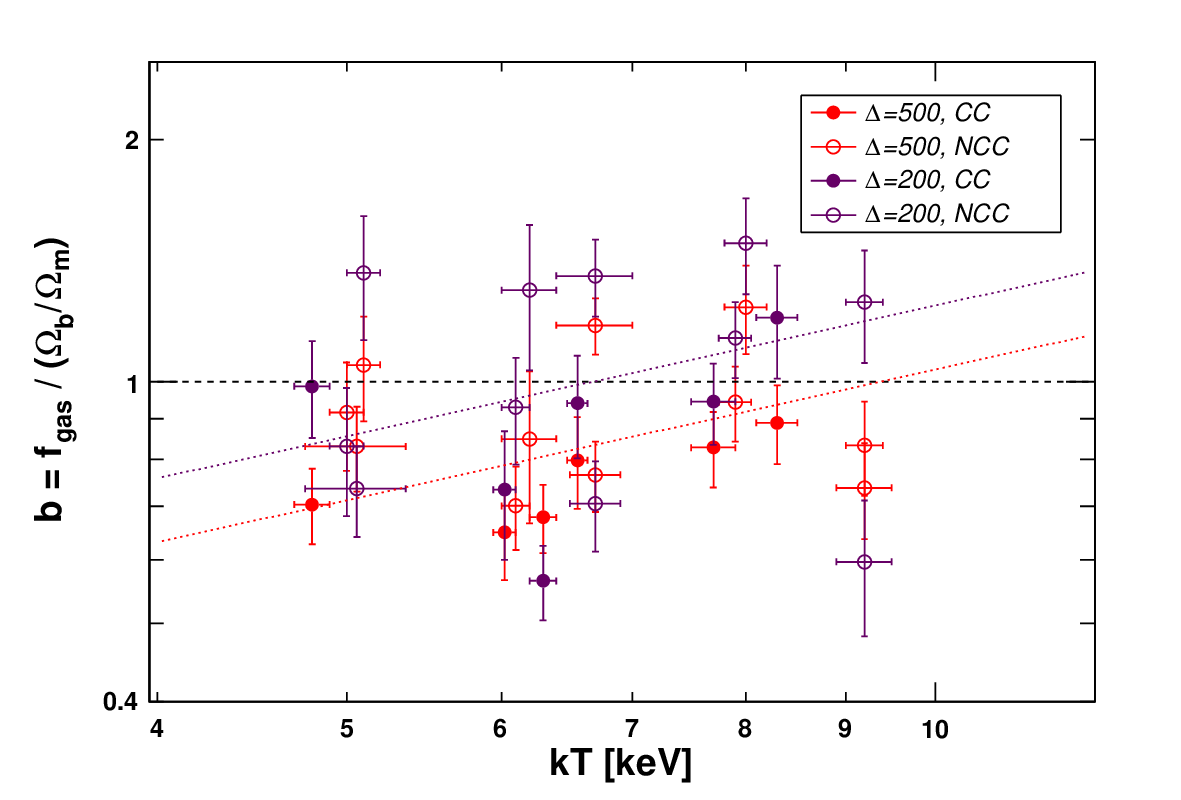}}}
\caption{Distribution of the gas mass fraction estimates normalized to the cosmic baryon fraction \citep{wmap7} as a function of the cluster's temperature (from E12) and core state at different overdensities. The error bars are given at the $1\sigma$ level. The dashed lines show the best fit to the entire sample using Eq. \ref{funcover}. For better readability, A2163 ($kT\sim18$ keV) is not shown in the plots}
\label{fgasTdelta}
\end{figure*}

In all cases, we can see that the enclosed gas fraction increases steadily with radius and gas temperature, as expected from previous work 
\citep[see e.g.,][]{arnaudlxt,ettori99,vikhlinin06}. To quantify the dependence on the gas temperature at different mass overdensities (200, 500, 1000, and 2500, see Fig.~\ref{fgasTdelta}), we fit a multi-linear function of the form
\begin{equation} \frac{f_{\rm gas}}{f_{\rm b, WMAP7}} = b_{{\rm gas },500}\left(\frac{\Delta}{500}\right)^\alpha \left(\frac{T_{vir}}{7\mbox{ keV}}\right)^\beta , \label{funcover}\end{equation}

\noindent where $b_{\rm gas}= f_{\rm gas}/f_{\rm b, WMAP7}$ is the depletion factor. The best-fitting values for the parameters are provided in Table \ref{tableover}. Only the normalization $b_{500}$ is found to change significantly between the CC and NCC populations, in agreement with Fig. \ref{fig:fgas}. The quality of the fit, however, is found to change significantly when comparing the CC and NCC populations, with only the fit to CC objects being formally acceptable. This indicates a significant cluster-to-cluster scatter in the latter population.

\begin{table}
\caption{\label{tableover}Best-fitting parameters for the function given in Eq. \ref{funcover}.}
\begin{center}
\begin{tabular}{lcccc}
\hline
Dataset & $b_{{\rm gas},500}$ & $\alpha$ & $\beta$ & $\chi^2$/d.o.f\\
\hline
\hline
Total & $0.854 \pm 0.016$ & $-0.20 \pm 0.02$ & $0.54 \pm 0.05$ & $154.4/69$ \\
CC & $0.757 \pm 0.024$& $-0.19 \pm 0.03$ & $0.52 \pm 0.14$ & $21.2/21$ \\
NCC & $0.919 \pm 0.021$ & $-0.20 \pm 0.02$ & $0.46 \pm 0.06$ & $100.7/45$\\
\hline
\end{tabular} 
\end{center}
\end{table}

\section{Discussion}

We discuss here two important implications of our results on (i) the use of gas mass fraction as cosmological probe and (ii) the comparison with predictions from hydrodynamically simulated objects.

\subsection{Implications on the use of clusters for cosmology}

The use of $f_{\rm gas}$ as a standard ruler for cosmology \citep{ettori03,allen08} implies that galaxy clusters are representative of the cosmic baryonic budget.
In the previous section, we quantified how the gas mass fraction depends on the gas temperature and matter overdensity, providing for the first time an observationally constrained calibration of the cluster baryon budget in the hot X-ray emitting phase. We assess here the cluster-to-cluster scatter of $f_{\rm gas}$ at different overdensity radii. To compute the intrinsic scatter of a given population, we used a maximum-likelihood estimator as introduced by \citet{maccacaro} (see also Appendix A of E12). This method allows the mean value and the intrinsic scatter to be computed jointly. In Table \ref{tabscat} we give the mean $f_{\rm gas}$ values and the intrinsic scatter $\sigma_f$ at overdensities of 2500, 1000, 500, and 200, for the entire population as well as for the CC and NCC clusters separately. We also report the scatter obtained when correcting the $f_{\rm gas}$ values for their dependence on the cluster's virial temperature, i.e., by rescaling each value by the quantity $(T_{vir}/\mbox{7 keV})^\beta$ (see Eq. \ref{funcover} and Table \ref{tableover}). 

In Table \ref{tabscat}, we can see that the level of scatter increases with radius, from $\sim15\%$ of the measured value at $R_{2500}$ to $\sim25\%$ at $R_{200}$. As expected, when a temperature scaling is performed, a lower scatter is measured in all cases. Interestingly, we note that the level of scatter is substantially higher for the NCC population than for the more relaxed, CC objects. When performing a temperature scaling, the level of intrinsic scatter is even consistent with 0 in CC systems out to $R_{500}$, while at $R_{200}$ some scatter at the level of $\sim18\%$ is measured, indicating that at the outermost radii even the most relaxed systems are affected by non-gravitational physics and/or inhomogeneous gas distributions (see E12). In the case of NCC systems, substantial cluster-to-cluster variations are found at all radii.

We remark that the trend of increasing scatter with radius is opposite to what we would expect when probing a larger volume, which shows that the the trend is likely caused by a physical effect. Both non-gravitational effects (turbulence, cosmic rays) and inhomogeneous gas densities can qualitatively explain this result. Indeed, numerical simulations show that non-thermal effects can be several times stronger in NCC systems \citep[e.g.,][]{lau09,pfrommer07,vazza11a,vazza12a}. Gas clumping and large-scale asymmetries are also expected to be more prominent along large-scale filaments \citep{nagai,vazza12c}, and thus the effect should be stronger in merging systems. For instance, if clumping biases the observed density high by $\sim20\%$, as found in Paper I for NCC clusters, the actual gas fraction for this class of objects would be consistent with the expected hot gas fraction. For a more detailed discussion of the implications of our results on the physics of cluster outskirts, we refer the reader to Paper I.

To summarize, our findings (both on the global baryon budget and the variations in $f_{\rm gas}$ from one system to another) demonstrate that the gas fraction of CC systems can be used efficiently as a standard ruler, while in NCC systems non-gravitational effects and/or inhomogeneities in the gas distribution introduce significant systematics in the computation of the gas fraction. A very similar conclusion was recently obtained by \citet{mahdavi12} by comparing X-ray and weak-lensing mass estimates. Indeed, X-ray mass estimates appear to be underestimated by $\sim15\%$ in NCC clusters, while good agreement is found for CC systems. The fact that a qualitatively similar result was obtained using different samples and techniques thus reinforces our conclusion. 

\subsection{Comparison with numerical simulations}

Using the shock-capturing Eulerian adaptive mesh refinement N-body+gas-dynamical adaptive refinement tree code, \citet{kravtsov05} measure in the most massive system of $M_{200} \approx 9.6 \times 10^{14} M_{\odot}$ a value of $b_{\rm gas}$ that rises from 0.86 (at $\Delta=2500$) to 1.04 (at the virial radius) in the run without dissipation and from 0.56 to 0.76 when radiative cooling, star formation, metal enrichment, and stellar feedback are considered. The constraints we obtain lie between these predictions, suggesting that both radiative processes are required and the feedback provided has to be less incisive in its action on distributing the ICM.

In a set of hydrodynamical simulations of massive galaxy clusters performed using the Tree+SPH code \texttt{GADGET-2}, \citet{ettori06} \citep[see also][]{deboni} estimate at $z=0$ a $b_{\rm gas}$ that increases from 0.82 to 0.89 from $\Delta=2500$ to $200$ in runs with gravitational heating only. On the other extreme, introducing cooling and star formation but no winds drastically reduces the expected relative gas fraction (0.29 at $\Delta=2500$; 0.59 at $\Delta=200$). Overall, however, none of the feedback models explored in that work are able to reproduce the high $b_{\rm gas}$ value of $0.74-0.9$ we observe at $\Delta < 500$, apart from the case where the alternative artificial low-viscosity scheme from \citet{morris} is considered, allowing about 30 per cent of the ICM thermal energy in the form of turbulent motion. 

For massive halos extracted from a non-radiative gas-dynamical realization of the {\it Millennium Simulations}, \citet{crain} measured $b_{\rm bar}$ of about 0.9 (and rms of 6 per cent) with almost no dependence on the radius between $R_{500}$ and $R_{200}$, implying a relative contribution from stars, with respect to the gas component, of 0.22 and 0.19 at the two radii, respectively, slightly higher than the present observational constraints \citep[e.g., $<0.17$ for $M_{500} > 3 \times 10^{14} M_{\odot}$ in][]{lagana}. In the {\it Millennium Gas Project}, \citet{millenniumgas} studied the baryon distribution in 170 objects, as SPH re-simulation of the Millennium simulation,  with gas temperature higher than 3 keV. They considered models where the ICM is heated (i) solely by gravitational process, (ii) by pre-heating in the order of 200 keV cm$^2$ occuring at $z=4$ combined with cooling, (iii) by feedback from supernovae (SNe) and active galactic nuclei (AGN) as expected from the semi-analytic predictions on galaxy formation but not including radiative cooling. For the most massive systems, they measured $b_{\rm gas}$ in the range 0.88--0.91, 0.57--0.83, 0.70--0.82 between $\Delta=2500$ and $200$ for cases (i), (ii), (iii), respectively, suggesting that (a) some cooling is needed to lower the gas fraction predicted in the inner part and (b) the implemented action of either pre-heating or feedback from SNe and AGN is probably too strong to fully explain the observed gas fraction in the clusters' outskirts.

\section{Conclusion}

Combining X-ray (\emph{ROSAT}/PSPC) and SZ (\emph{Planck}) data, we analyzed the properties of the ICM at large radii using both sample-averaged profiles and the data for 18 individual systems. This allowed us to measure for the first time hydrostatic mass and gas fraction profiles out to the virial radius in a relatively large sample of objects. The results on the thermodynamic properties of galaxy clusters close to the virial radius are presented in a companion paper. Our results can be summarized as follows:

\begin{itemize}
\item
The gas fraction reconstructed through the hydrostatic equilibrium equation increases with radius and reaches the value of $f_{\rm gas,200}=0.176\pm0.009$ around $R_{200}$. Different input data and deprojection techniques yield results that excellently agree. This value is consistent with the cosmic baryon fraction $\Omega_b/\Omega_m=0.167$ \citep{wmap7}. Given that the stellar content is expected to account for $10-20\%$ of the baryons in galaxy clusters, the measured gas fraction slightly exceeds the expected value. A 15\% bias  either in raising the hydrostatic masses, as expected from numerical simulations, or in lowering the gas mass allows us to reconcile our measurement with the expected gas fraction.

\item We observe differences between the gas fraction measured in relaxed, CC systems compared to dynamically active, NCC systems. In NCC clusters, the gas fraction exceeds the cosmic baryon fraction significantly, indicating that hydrostatic equilibrium breaks down in the outer regions of perturbed systems, and/or that the gas is distributed in an inhomogeneous way. Conversely, in CC systems the gas fraction converges to the expected hot gas fraction, confirming that they correspond to more relaxed systems, which can be used more efficiently for cosmological purposes.

\item From the 18 objects comprising our sample, we measured the scatter around the average $f_{\rm gas}$ value using a maximum-likelihood estimator. We find that the scatter increases with radius, from 15\% at $R_{2500}$ to 25\% at $R_{200}$. In contrast to NCC systems, in which we measure substantial cluster-to-cluster variations at all radii, CC systems exhibit a negligible level of scatter out to $R_{500}$, confirming that they are better suited for cosmological studies.

\item We quantified the $f_{\rm gas}$ dependence upon gas temperature and matter overdensity with the relation $f_{\rm gas}/f_{\rm b, WMAP7} = b_{500} \left(\Delta/500\right)^\alpha \left(T_{\rm gas}/7\mbox{ keV}\right)^\beta$ and measured $b_{500} = 0.76 \pm 0.02$ in CC systems and $0.92 \pm 0.02$ in NCC clusters, $\alpha=-0.2$ and $\beta=0.5$. The gas bias (or depletion) factor $b_{500}$, and its variation with temperature and radii, disagrees with the results obtained for clusters extracted from recent hydrodynamical simulations, implying that the action of cooling and AGN and/or SN feedback is indeed needed but has to be regulated differently from what has been used so far.
\end{itemize}

\bibliographystyle{aa}
\bibliography{aa20403}

\end{document}